# NEW OBSERVATIONS REGARDING DETERMINISTIC, TIME REVERSIBLE THERMOSTATS AND GAUSS'S PRINCIPLE OF LEAST CONSTRAINT

Joanne N. Bright* and Denis J. Evans,

Research School of Chemistry, Australian National University, Canberra, ACT

0200, AUSTRALIA

Debra J. Searles,

School of Science, Griffith University, Brisbane, Qld 4111, AUSTRALIA

*Deterministic thermostats are frequently employed in non-equilibrium molecular dynamics simulations in order to remove the heat produced irreversibly over the course of such simulations. The simplest thermostat is the Gaussian thermostat, which satisfies Gauss's principle of least constraint and fixes the peculiar kinetic energy. There are of course infinitely many ways to thermostat systems, e.g. by fixing $\sum_i |p_i|^{m+1}$. In the present paper we provide, for the first time, convincing arguments as to why the conventional Gaussian isokinetic thermostat ($m=1$) is unique in this class. We show that this thermostat minimizes the phase space compression and is the only thermostat for which the conjugate pairing rule (CPR) holds. Moreover it is shown that for finite sized systems in the absence of an applied dissipative field, all other thermostats ($m \neq 1$) perform work on the system in the same manner as a dissipative field while simultaneously removing the dissipative heat so generated. All other thermostats ($m \neq 1$) are thus auto-dissipative. Among all $m$-thermostats, only the $m=1$ Gaussian thermostat permits an equilibrium state.*

**Introduction**

In 1829 Carl Friedrich Gauss established the dynamical principle now known as Gauss's Principle of Least Constraint,[1] stating that a system subject to constraints will follow trajectories which, in a least-squares sense, differ minimally from their unconstrained Newtonian counterparts. The principle applies to all constraints whether they are holonomic (involving constraints that depend only on coordinates), or non-holonomic (which involve non-integrable constraints on velocity). Gauss's principle was employed independently by Hoover et al.[2] and Evans[3] to develop time reversible deterministic thermostats for molecular dynamics computer simulations. In particular, the heat produced irreversibly by an external field can be removed from the system by simple modifications to the equations of motion in the form of thermostatting constraints.[4,5]

In a real physical system heat is removed by conduction, radiation or convection to the boundaries. The process can be represented explicitly by modelling isothermal reservoirs surrounding the system of interest. The reservoirs exchange heat with the system via interparticle interactions. Gaussian thermostats avoid the need to model these complex system-reservoir interactions. They also minimize system size dependence and simulation time. The effect of the reservoirs is thus reproduced in a simple manner that can be employed in non-equilibrium simulations to allow for the possibility and maintenance of a steady state.

3The removal of heat by thermostatting forces leads to volumes in phase space being no longer preserved i.e. a reduction in the volume of accessible phase space or phase space compression.[6] For real, macroscopic systems this phase space compression and the associated dimension loss are insignificantly small[7] and the system evolves to a strange attractor of similar dimensionality to the unperturbed system. In small systems the reduction can be more pronounced.

**Gauss's principle of least constraint**

For a system described by coordinates, $\mathbf{r} \equiv (\mathbf{r}_1, \mathbf{r}_2, ...)$, and time, t, constraints confine trajectories to a hypersurface (the constraint plane), defined by $g(\mathbf{r}, \dot{\mathbf{r}}, t) = 0$. Differentiating $g$ with respect to time results in the differential constraint equation, which imposes a condition on the acceleration vector of particles within the system:[8]

$$\boldsymbol{\beta}(\mathbf{r}, \dot{\mathbf{r}}, t) \bullet \ddot{\mathbf{r}} + g(\mathbf{r}, \dot{\mathbf{r}}, t) = 0 \tag{1}$$

where $\beta$ and $\gamma$ are:[9]

$$\begin{aligned} \boldsymbol{\beta} &= \frac{\partial g}{\partial \dot{\mathbf{r}}} \\ g &= \dot{\mathbf{r}} \bullet \frac{\partial g}{\partial \mathbf{r}} + \frac{\partial g}{\partial t} \end{aligned} \tag{2}$$

While unconstrained trajectories obeying Newtonian equations of motion, $\ddot{\mathbf{r}}_u = \mathbf{F}/m$, are free to leave the constraint plane, constrained trajectories following the equations of motion, $\ddot{\mathbf{r}} = \mathbf{F}/m - x\boldsymbol{\beta}/m$, are prevented from doing so by the application of the additional constraint "force", $-x\boldsymbol{\beta}$, satisfying Eqs (1) and (2):

$$x = \frac{(\boldsymbol{\beta} \bullet \ddot{\mathbf{r}}_u + \gamma)}{\boldsymbol{\beta} \bullet \boldsymbol{\beta}/m}. \qquad (3)$$

When multiple constraints are imposed each constraint forces is added and the constraint multipliers, which may be coupled, are then determined. For a simple Gaussian thermostat which fixes the kinetic energy of the system such that $g(\mathbf{r},\dot{\mathbf{r}},t) = \frac{m\dot{\mathbf{r}}^2}{2} - E_{kin} = 0$, the constrained equations of motion are $\ddot{\mathbf{r}} = \mathbf{F}/m - x\dot{\mathbf{r}}$, and it can easily be shown that $x = \frac{\mathbf{F} \cdot \dot{\mathbf{r}}}{m\dot{\mathbf{r}}^2}$.

It is important to note that there is no unique means of projecting the unconstrained trajectories onto the constant hypersurface and although Gauss's principle defines one method, a multiplicity of methods may be employed. We discuss such methods in what follows.

## $m$-Thermostats in Equilibrium and Non-equilibrium Systems

The properties of a number of "$m$" thermostats and ergostats (which fix the internal energy of the system) have been explored in the weak field regime.[10,11] In these papers the system was described by the equations of motion:

$$\dot{q}_{id} = \frac{p_{id}}{m} + C_{idg} F_{eg}$$
$$\dot{p}_{id} = F_{id} + D_{idg} F_{eg} - a |p_{id}|^{m-1} p_{id} \quad (4)$$

where Einstein notation is used, $d, g = x, y, z$, $q_{id}$ is the position of the $i$-th particle in the $d$-direction, $p_{id}$ is the momentum of the $i$th particle in the $d$-direction, $C_{idg}$ and $D_{idg}$ couple the system with the external field, $F_{eg}$, and

$$a = \frac{\sum_{i=1}^{N} \sum_{d,g=x,y,z} (p_{id} D_{idg}/m - F_{ig} C_{idg}) F_{eg}}{\sum_{i=1}^{N} \sum_{d=x,y,z} |p_{id}|^{m+1}/m}$$

$$\equiv \frac{-\mathbf{J} \bullet \mathbf{F}_e V}{\sum_{i=1}^{N} \sum_{d=x,y,z} |p_{id}|^{m+1}/m} \quad (5)$$

is the ergostat multiplier that was used to fix the internal energy as a constant. $\mathbf{J}$ is the (intensive) dissipative flux and is related to the unthermostatted (adiabatic) rate of change of the internal energy $H_0$ via $\left.\frac{dH_0}{dt}\right|^{ad} = -\mathbf{J}(\Gamma)V \bullet \mathbf{F}_e$. Note that in the absence of the perturbing field and when the internal energy of the system is held constant, $\boldsymbol{a}$ is obviously zero and hence does no work on the system at equilibrium. It was shown that for a range of perturbing fields and different values of $\boldsymbol{m}$, the Gaussian ergostat (for which $\boldsymbol{m}=1$) minimized the magnitude of the change in acceleration brought about by the constraint.[10] The average value of the phase space compression factor $\Lambda$ (equivalent to the logarithmic time rate of change of the N-particle distribution function) was also studied for these systems and it was shown to be minimal for a Gaussian ergostatted system.[10]

In this paper we focus on the isokinetic case. Since the kinetic energy is not a constant in the unthermostatted (adiabatic) system, the isokinetic thermostatting multiplier, $\boldsymbol{a}$, is not instantaneously zero at equilibrium. We employ a series of "$\boldsymbol{m}$" thermostats to fix either the second moment of the momentum distribution (or equivalently fix the kinetic energy $K_2$), or the "($\boldsymbol{m}+1$)th" moment of the velocities, $K_{\boldsymbol{m}+1} = \sum_{i=1,\boldsymbol{d}}^{N} \frac{|p_{i\boldsymbol{d}}|^{\boldsymbol{m}+1}}{2m}$ and consider systems at equilibrium and also those under the influence of a weak "colour" field $\mathbf{F}_e$.[8] The $\boldsymbol{m}$-thermostatted equations of motion for the $i$-th particle in this system are:

$$\dot{q}_{i\boldsymbol{d}} = \frac{p_{i\boldsymbol{d}}}{m}$$
$$\dot{p}_{i\boldsymbol{d}} = F_{i\boldsymbol{d}} + c_i F_{e\boldsymbol{d}} - \boldsymbol{a}|p_{i\boldsymbol{d}}|^{\boldsymbol{m}-1} p_{i\boldsymbol{d}} + z_{\boldsymbol{d}}$$
(6)

where $c_i = (-1)^i$ and denotes the 'colour' of the $i$-th particle, $z_d = a \frac{\sum_{i=1}^{N} p_{id} |p_{id}|^{m-1}}{N}$ is a Gaussian constraint introduced to keep the momentum in the $\delta$ direction fixed. When $K_2$ is fixed, the thermostatting variable $a$ can easily be determined to be:

$$a_{K_2} = \frac{\sum_{i,d} F_{id} p_{id} + F_{ed} c_i p_{id}}{\sum_{i,d} |p_{id}|^{m+1}} \quad . \tag{7}$$

Similarly, fixing the ($m+1$)th moment, $K_{m+1}$ with a $m$-thermostat, gives:

$$a_{K_{m+1}} = \frac{\sum_{i,d} F_{id} p_{id} |p_{id}|^{m-1} + F_{ed} c_i p_{id} |p_{id}|^{m-1}}{\sum_{i,d} \left( p_{id}^{2m} \right) - \sum_{d} \left( \sum_{i} |p_{id}|^{m-1} p_{id} \right)^2 / N} \quad . \tag{8}$$

Clearly, setting $m$ equal to 1 in either case returns the usual Gaussian thermostat. Note however that in the first case, where $K_2$ is fixed and $m$ is varied, Gauss's principle of least constraint is only satisfied for the thermostatted constraint when $m=1$. In the latter case, where $K_{m+1}$ is fixed, Gauss's principle is obeyed for the themostatting and total momentum constraints for all values of $m$. Those constraints that satisfy Gauss's principle are constrained via the least possible change in acceleration (most direct path) to the constraint surface, whereas those that violate it may require greater perturbation from the unconstrained trajectories.

The systems we consider in this paper all satisfy the condition that, in the absence of the thermostat, they preserve volumes in phase space i.e. adiabatic incompressibility of phase space (AIΓ).[8] The presence of a thermostat, however, leads to the possibility of phase space compression that is quantified by the phase space compression factor, $\Lambda$. The phase space compression factor is the rate of change of the logarithm of the N-particle distribution function $f(\Gamma,t)$, where $\Gamma \equiv (\mathbf{q},\mathbf{p})$:

$$\Lambda(\Gamma) = -\frac{d \ln f(\Gamma,t)}{dt} = \frac{\partial}{\partial \Gamma} \cdot \dot{\Gamma}, \tag{9}$$

and describes the reduction in the available phase space to the system. For an equilibrium systems where a $m$-thermostat fixes $K_2$, $\Lambda$ is:

$$\Lambda(\Gamma) = -\alpha m(1-1/N)\sum_{id}\left|p_{id}\right|^{m-1} + A + B \tag{10}$$

where

$$A = -\frac{\sum\limits_{i=1,d}^{N} F_{id}\left(\left|p_{id}\right|^{m-1} p_{id} - (1/N)\sum\limits_{j=1}^{N}\left|p_{jd}\right|^{m-1} p_{jd}\right)}{\sum\limits_{i=1,d}^{N}\left|p_{id}\right|^{m+1}}$$

and

$$B = \alpha(m+1)\frac{\sum\limits_{i=1,d}^{N}\left(p_{id}^{2m}\right) - (1/N)\sum\limits_{d}\left(\sum\limits_{i=1}^{N}\left|p_{id}\right|^{m-1} p_{id}\right)^{2}}{\sum\limits_{i=1,d}^{N}\left|p_{id}\right|^{m+1}}$$

A similar expression can be derived for Gaussian thermostats that fix $K_{m+1}$:

$$\Lambda(\Gamma) = -am(1-1/N)\sum_{id}|p_{id}|^{m-1} + C + D \quad (11)$$

where

$$C = -m\frac{\sum_{i=1,d}^{N} F_{id}\left(p_{id}^{2m-1} - (1/N)|p_{id}|^{m-1}\sum_{j}^{N}|p_{jd}|^{m-1} p_{jd}\right)}{\sum_{i,d}\left(p_{id}^{2m}\right) - \sum_{d}\left(\sum_{i}|p_{id}|^{m-1} p_{id}\right)^2 / N}$$

and

$$D = 2ma\frac{\sum_{i=1,d}^{N}\left(|p_{id}|^{3m-1} - (2/N)p_{id}^{2m-1}\sum_{j=1}^{N}\left(|p_{jd}|^{m-1} p_{jd}\right) + (1/N^2)p_{id}^{m-1}\left(\sum_{j=1}^{N}|p_{jd}|^{m-1} p_{jd}\right)^2\right)}{\sum_{i,d}\left(p_{id}^{2m}\right) - \sum_{d}\left(\sum_{i}|p_{id}|^{m-1} p_{id}\right)^2 / N}$$

.

In both systems $\Lambda$ can be written:

$$\Lambda(\Gamma) = -am\sum_{id}|p_{id}|^{m-1} + O_N(1) \quad (12)$$

and is related by a simple equation to the rate of change of the fine grained Gibbs entropy

$S(t) \equiv -k_B \int d\Gamma f(\Gamma,t) \ln f(\Gamma,t)$ i.e. $\frac{dS(t)}{dt} = k_B \langle \Lambda(t) \rangle$. For the case of a standard Gaussian isokinetic thermostat ($m=1$), equation (10) shows that $\Lambda$ is simply given by $\Lambda = -3N\mathbf{a}(1+c/N))$ where $c$ is a constant. Since in this case $\mathbf{a}$ can be written

$\mathbf{a} = \frac{-\dot{H}_0}{2mK_2}$ where $K_2$ is a constant it is clear that if a steady state is reached then

$\langle \dot{H}_0 \rangle = 0$ and therefore $\langle \mathbf{a} \rangle = 0$ and $\langle \Lambda \rangle = 0$ for all $N$. The question, however, of whether the same holds true for isokinetic ($K_2$) systems and for Gaussian iso-$K_{m+1}$ systems with $m \neq 1$ remains untested.

The dynamic behaviour of a system can also be described in terms of its Lyapunov exponents, which measure the exponential rate of divergence of nearby trajectories in phase space. If we define a displacement vector $\Delta\Gamma_j(t) = \Gamma_j(t) - \Gamma(t)$ measuring separation between nearby points $\Gamma_j(t)$ and $\Gamma(t)$ in phase space then, in the limit of small displacements, the vectors become tangent vectors obeying equations of motion: $d\dot{\Gamma}_i = T \cdot d\Gamma_i$ where $T = \partial\dot{\Gamma}/\partial\Gamma$ is the Jacobian or stability matrix of the flow. The maximum Lyapunov exponent, which is only defined in the long time limit, is given by $l_1 = \lim_{t \to \infty} \frac{1}{t} \ln \frac{|d\Gamma_1(t)|}{|d\Gamma_1(0)|}$.[8,12] This describes the asymptotic rate of exponential separation of two nearby points in phase space. Consider a set of tangent vectors $\{d\Gamma_i; i=1,2,...2dN\}$ that evolve according to the equations of motion, but are constrained to remain orthogonal to each other so that $d\Gamma_i$ is orthogonal to vectors $\{d\Gamma_j; j<i\}$. The value of d is the number of Cartesian dimensions considered and 2dN is therefore the dimension of phase space. This set of orthogonal tangent vectors will give the full set of 2dN Lyapunov exponents, defined by $l_i = \lim_{t\to\infty} \frac{1}{t} \ln \frac{|d\Gamma_i(t)|}{|d\Gamma_i(0)|}$. Phase volumes defined by these tangent vectors will grow exponentially at a rate given by the sum of the corresponding Lyapunov exponents, and the time evolution of an infinitesimal volume in the full phase space is given by the total sum of all the Lyapunov exponents. It is related to the phase space compression by the simple relation: $\Lambda = \sum_{i=1}^{2dN} l_i$.

The Conjugate Pairing Rule (CPR) states that the sum of conjugate pairs of Lyapunov exponents ($l_i + l_{2dN+1-i}, \forall i$) is zero at equilibrium[12] and equal to a constant,[13] independent of the pair index, in a field driven system. For systems satisfying the CPR, it is possible to calculate transport coefficients and entropy production from the maximal ($i=1$ and $i=2dN$) exponents alone.[13] Systems that are symplectic in the absence of thermostats (this includes all Hamiltonian systems), and are thermostatted homogeneously by a Gaussian isokinetic thermostat, satisfy CPR - apart from certain zero exponents.[14,15]

In previous work[10] it was shown that CPR is violated in non-equilibrium isoenergetic $m$-thermostatted systems ($m \neq 1$). Equilibrium and non-equilibrium isokinetic systems however were not explored. We discuss the Lyapunov spectra and adherence to CPR in such systems here.

## Results

To explore the behaviour of *m*-thermostatted systems when different moments of the kinetic energy are fixed, we simulated a number of 2D, soft disc systems both at equilibrium (zero field) and in the presence of a weak "colour" field $\mathbf{F}_e = (F_{ex}, 0)$. The particles interact via a short range WCA potential and standard periodic boundary conditions were employed. The equations of motion were integrated using a fourth order Runge-Kutta integration scheme with a time step of 0.0005 (all units are reduced Lennard-Jones units). The temperature in the simulations was fixed at $T = 1.0$ (or alternatively, a temperature of $T$ was established at the beginning of the simulation and the resulting value of $K_{m+1}$ held constant), and the number of particles $N$ was either 4 or 50. Two reduced densities, 0.4 and 0.8, were simulated and *m* was varied between 0.1 and 6.

## Comparison of Thermostats in Equilibrium Systems

Figures 1 and 2 illustrate the behaviour of a typical system, in the absence of field, for which the second moment of the momentum (or $K_2$), has been fixed with varying values of *m*. It is clear that both *a* and $-\Lambda$ are minimised, and are equal to zero for the standard Gaussian *p* thermostat (*m*=1). The fact that both *a* and $\Lambda$ are non-zero for all other values of *m* is indicative of the importance of adherence to Gauss's principle even at equilibrium. The thermostat in these cases (µ≠1) does work on the system, driving it away from equilibrium in the same manner as a dissipative field but at the same time extracts the dissipative heat so generated so as to generate a non-equilibrium steady state (rather than a true equilibrium state).

Insert Figure 1 near here

Insert Figure 2 near here.

This result can be described with reference to Gauss's principle of least constraint. In fixing the second moment of the temperature, $K_2$, with a $p$ ($m=1$) thermostat, the least change in acceleration is applied to trajectories resulting in minimal deviation from the unconstrained trajectory paths. In terms of the phase space, this corresponds to the shortest projection path of the unconstrained system to the constraint plane. In this case, the time average of fluctuations in the constraint force and the phase space compression are both zero if the system is at equilibrium.

For all other values of $m$ the constraint force is no longer minimal; work is done on the system by the thermostat and the phase space is compressed. This is an important result as it indicates the importance of thermostat constraints that adhere to Gauss's principle even at equilibrium. Indeed in the absence of an explicitly applied external field $\mathbf{F}_e = \mathbf{0}$, it is only a $m=1$ Gaussian isokinetic thermostat that generates an equilibrium state. All other $m$ thermostats are *auto-dissipative* and possess no equilibrium state.

It is also interesting to consider the properties of a $m$-thermostatted system in which $K_{m+1} = \sum_{i,d} |p_{id}|^{m+1} / 2m$ is constrained such that Gauss's principle is satisfied for $m \neq 1$.

For example, $m=3$ constrains the fourth moment of the momentum: $K_4 = \sum_{i,d} |p_{id}|^4 / 2m$

via the thermostatting variable $a_{K_4} = \dfrac{\sum_{i,d} F_{id} p_{id}^3}{\sum_{i,d} p_{id}^6 - \sum_{d} \left( \sum_{i} p_{id}^3 \right)^2 / N}$.

The results for a typical system over a range of values of $m$ are plotted in Figures 3 and 4. Clearly in this case the average value of the thermostatting variable is zero, independent of the value of $m$. This is to be expected since in this case $\dot{H}_0 = -2mK_{m+1}a$, where we have used the equations of motion given by equation (6) and the fact that the total momentum is zero. If the system is to reach a steady state, $\langle \dot{H}_0 \rangle = 0$ and since $K_{m+1}$ is held constant, $\langle a \rangle = 0$ for all $m$ and all $N$. However, this does **not** imply that $\langle \Lambda \rangle = 0$ unless $m=1$ (discussed above), since $\Lambda$ and $a$ are not directly proportional when $m \neq 1$ (see equation (11)).

Insert Figure 3 near here.

Insert Figure 4 near here.

While the $K_{m+1}$ thermostat applies the minimum change in acceleration in projecting the trajectories onto the constraint plane (note that this plane differs depending on the value of *m* and hence the particular constrained moment), it acts like a dissipative field, attempting to change the shape of the velocity distribution function while simultaneously removing the dissipative heat generated by this attempt to deform the shape of the velocity distribution function. This is evident in Figure 4 where the phase space compression shows a similar behaviour to the *m*-thermostatted systems which do not satisfy Gauss's principle, exhibiting a clear minimum for the case *m*=1 constraining $K_2$. Referring to Equation (12), while *a* averages to zero, correlations between its fluctuations with those of $\sum_{i=1,d}^{N} |p_{id}|^{m-1}$ lead to non-zero phase space compression $\Lambda$. Thus while Gauss's principle holds true at every phase point, on average the overall phase space contracts.

In order to understand the behaviour of these systems, it is important to ask what happens to the iso-$K_\chi$ N-particle distribution function:

$$f_{K_c}(\Gamma) = \frac{\exp(-b\Phi(\Gamma))\delta(K_c(\Gamma)-K_{c,0})}{\int d\Gamma \exp(-b\Phi(\Gamma))\delta(K_c(\Gamma)-K_{c,0})}$$ over time. For systems with finite N, we consider whether, as has been previously suggested,[8] the isokinetic distribution function is preserved by *m*-thermostatted dynamics at equilibrium ? (i.e. does $\frac{\partial f_{Kc}}{\partial t} = 0$ when $\mathbf{F}_e = \mathbf{0}$ ? )

$$\frac{\partial f_{Kc}}{\partial t} = -\dot{\Gamma} \bullet \frac{\partial f_{Kc}}{\partial \Gamma} - f_{Kc} \frac{\partial}{\partial \Gamma} \bullet \dot{\Gamma}$$

$$= \left( b\dot{\Phi} + ma \sum_{i=1,d}^{N} |p_{id}|^{m-1} + O_N(1) \right) f_{Kc} \quad (13)$$

Consider first the case where $c = 2$, i.e. the second moment of the momentum is constrained. In this case,

$$\frac{\partial f_{K_2}}{\partial t} = \left( -b \sum_{i=1,d}^{N} F_{id} p_{id} + m \sum_{i=1,d}^{N} F_{id} p_{id} \frac{\sum_{i=1,d}^{N} |p_{id}|^{m-1}}{\sum_{i=1,d}^{N} |p_{id}|^{m+1}} + O_N(1) \right) f_{K2} \quad (14)$$

For the standard Gaussian thermostat, $m = 1$, this reduces to:

$$\frac{\partial f_{K_2}}{\partial t} = (-b \sum_{i=1,d}^{N} F_{id} p_{id} + \frac{(dN-d-1)}{K_2} \sum_{i=1,d}^{N} F_{id} p_{id}) f_{K2}$$

$$= 0 \quad \text{iff} \quad b = \frac{(dN-d-1)}{K_2} \quad (15)$$

Clearly for $m \neq 1$ the distribution function is preserved only when

$$\sum_{i=1,d}^{N} |p_{id}|^{m-1} = \frac{b}{m} \sum_{i=1,d}^{N} |p_{id}|^{m+1} \quad \text{or} \quad b = \frac{m \sum_{i=1,d}^{N} |p_{id}|^{m-1}}{2mK_{m+1}}.$$ This will only be true in the thermodynamic limit since for finite $N$, $K_{m+1}$ fluctuates ($K_2$ is constant). For the case $c = m+1$, a similar result can be derived i.e.

$$\frac{\P f_{Km+1}}{\P t} = \left( -b \sum_{i=1,d}^{N} F_{id} p_{id} + m \frac{\sum_{i=1,d}^{N} F_{id} p_{id} |p_{id}|^{m-1} \sum_{id} |p_{id}|^{m-1}}{\sum_{i,d}(p_{id}^{2m}) - \sum_{d}\left(\sum_{i} |p_{id}|^{m-1} p_{id}\right)^2 / N} + O_N(1) \right) f_{Km+1} \quad (16)$$

It is clear that for $\mu \neq 1$ there is no $\beta$ for which this distribution is preserved. However, in the thermodynamic limit where $b = m \dfrac{\sum_{i=1,d}^{N} F_{id} p_{id} |p_{id}|^{m-1} \sum_{id} |p_{id}|^{m-1}}{\sum_{i=1,d}^{N} F_{id} p_{id} (\sum_{i,d}(p_{id}^{2m}) - \sum_{d}(\sum_{i} |p_{id}|^{m-1} p_{id})^2 / N)}$ the

iso-$K_{\mu+1}$ distribution is preserved.

We can interpret the action of the **m** thermostat in both cases as an additional external field on the system superimposed over a regular Gaussian *p* thermostat i.e. if we rewrite the equation of motion for the momentum in the equilibrium system as:

$$\dot{p}_{id} = F_{id} - a\, p_{id} - a\left( \frac{p_{id}}{|p_{id}|} |p_{id}|^m - p_{id} \right) + z_d, \quad (17)$$

then we can identify the term $a_m = a\left( \dfrac{p_{id}}{|p_{id}|} |p_{id}|^m - p_{id} \right)$ with a dissipative field which attempts to change the shape of the distribution function. In the process the phase space volume is not preserved by the dynamics, resulting in a constant decrease of the Gibbs entropy of the system and constant compression of the occupied phase space. The new distribution function thus evolves to a strange attractor, possessing a lower dimensionality than the equilibrium distribution function of the regular Gaussian ($m=1$) thermostatted system.

**Influence of System Size on Thermostat Properties**

All the results above relate to a small system in which the distribution function at equilibrium is not conserved by $m \neq 1$ dynamics. It is interesting to compare the behaviour of both $a$ and $\Lambda$ in small and large system limits. Figures 5 and 6 plot the behaviour of these variables for $m$-thermostatted systems fixing $K_2$ containing 4 and 50 particles respectively. Similar results were obtained for $m$-thermostatted systems fixing $K_{m+1}$.

Insert Figure 5 near here

Insert Figure 6 near here

For the $K_2$ thermostatted systems, $\dot{H}_0 = -\sum_{i,d} F_{id} p_{id} / m$ and therefore it is clear that $\left\langle \sum_{i,d} F_{id} p_{id} \right\rangle = 0$ for any system that is at equilibrium or is in a steady state. If there is no explicit field applied, then in the thermodynamic limit,

$$a_{K_2} \equiv \left\langle a_{K_2} \right\rangle = \left\langle \sum_{i,d} F_{id} p_{id} \right\rangle \Big/ \left\langle \sum_{i,d} |p_{id}|^{m+1} \right\rangle = 0 \text{ (i.e. there are no fluctuations in } a_{K_2}\text{),}$$

consistent with the numerical results. The same conclusion can be drawn by noting that in the thermodynamic limit, $f_{K_2}(\Gamma) = \dfrac{\exp(-\boldsymbol{b}\Phi(\Gamma))\boldsymbol{d}(K_2(\Gamma)-K_{2,0})}{\int d\Gamma \exp(-\boldsymbol{b}\Phi(\Gamma))\boldsymbol{d}(K_2(\Gamma)-K_{2,0})}$, and that this distribution is even with respect to transformation of the coordinates $\mathbf{q} \to -\mathbf{q}$. Since $\boldsymbol{a}_{K_2}$ is odd with respect to this transformation, $\boldsymbol{a}_{K_2} \equiv \langle \boldsymbol{a}_{K_2} \rangle = 0$. A similar argument can be used to show $\boldsymbol{a}_{K_{m+1}} = \langle \boldsymbol{a}_{K_{m+1}} \rangle = 0$, and that in field free $K_2$ or $K_{m+1}$ thermostatted systems, $\Lambda \equiv \langle \Lambda \rangle = 0$.

Note that the minimum in both $\boldsymbol{a}$ and $-\Lambda$ become less pronounced as the system size is increased, confirming the theoretical results indicating that in the thermodynamic limit these variables average to zero.

**Lyapunov Spectra and The Conjugate Pairing Rule**

The Lyapunov exponents may be calculated numerically via several schemes discussed in detail previously.[15] The calculations presented here correspond to a method in which the equations of motion of a mother trajectory and an additional 2dN daughter trajectories (generated via infinitesimal displacements to the mother) are simulated and constrained to remain orthogonal to and a fixed phase space distance from mother. The Lyapunov exponents are obtained from the distance constraint multiplier as discussed in reference 14. We confirmed our results via alternative calculation methods.

In Figures 7 and 8 we plot Lyapunov spectra for several $m$-thermostatted $\mathbf{F}_e = \mathbf{0}$ systems.

Insert Figure 7 near here

Insert Figure 8 near here

Clearly for both $m$ thermostats fixing $K_2$ and $K_{m+1}$ the only value of $m$ for which the conjugate pairing rule is satisfied is $m=1$ i.e. the standard Gaussian thermostat. Other values of $m$ shift the spectrum to more negative values. In the light of Figs 2 and 4 this result is unsurprising. As the phase space is compressed the rate of contraction of infinitesimal areas in phase space dominates the rate of expansion with a resulting shift in the spectrum. The changes are most prominent in the smallest exponents and indicate evolution towards a strange attractor. We can estimate the dimension of this attractor by calculating the Kaplan-Yorke dimension of the $m$-thermostatted systems. The Kaplan-Yorke dimension,[16] $D_{KY}$, is given by:

$$D_{KY} = N_{KY} + \frac{\sum_{i=1}^{N_{KY}} l_i}{\left| l_{N_{KY}+1} \right|} \qquad (18)$$

where $N_{KY}$ is the largest integer for which $\sum_{i=1}^{N_{KY}} l_i$ is positive. Volume elements associated with Lyapunov exponents $i > D_{KY}$ contract in time.[16,17] For a system with $m=1$, $D_{KY} = 16$ i.e. no phase space contraction occurs. In contrast, for a system with $m=5$, such as illustrated in Figures 7 and 8, $D_{KY} = 14.8$ and 15.4 for $m$-thermostats fixing $K_2$ and $K_{m+1}$ respectively, indicative of the phase space contraction in these systems. The behaviour of Lyapunov spectra, the Kaplan-Yorke dimension and satisfaction of CPR with varying values of $m$ and varying perturbing fields will be examined in more detail in a forthcoming paper.

**Influence of Weak Fields on Thermostat Properties in Small Systems**

We also examined the effect of a weak colour field on systems thermostatted via the family of $m$-thermostats discussed above. The external field does work on the system that is then converted into heat that must be removed by the action of the thermostat. Equation (17) suggests that the dissipative action of a $m$-thermostat (i.e. $a_m$) and its thermostatting action in response to an external field can be treated independently. We can test this by comparing a series of $m$-thermostatted systems at equilibrium with those under the action of finite fields.

Figures 9 and 10 demonstrate that the influence of a weak colour field (applied here in the x direction) does little to alter the results presented above. As shown in Figures 9 and 10 the degree to which $a$ and $\Lambda$ change with $m$ changes little with the superimposed field i.e. a weak field produces a simple shift and does not change the relative behaviour of the thermostats.

**Uniqueness of the Gaussian $m = 1$ thermostat – momentum rescaling**

It is interesting to consider how $m$-thermostatting alters the momentum distribution, and thus how it might be expected to change the Lyapunov spectrum of the system. It is straightforward to show that continual, uniform rescaling of the momentum of each particle produces the same dynamics as a $m = 1$ thermostat. Using the finite difference relation to determine the time evolution of the unthermostatted (adiabatic) equations of motion for a system gives $\lim_{dt \to 0} p_{id}^{ad}(t + dt) = p_{id}(t) + \left( F_{id}(t) + D_{idg}(t) F_{eg} \right) dt$. The time evolved momentum in the thermostatted system is then

$$\begin{aligned} \lim_{dt \to 0} p_{id}(t + dt) &= p_{id}(t) + \left( F_{id}(t) + D_{idg}(t) F_{eg} \right) dt - \mathbf{a}(t) p_{id}(t) dt \\ &= p_{id}^{ad}(t + dt) - \mathbf{a}(t) p_{id}^{ad}(t + dt) dt + O(dt^2) \\ &= \left( 1 - \mathbf{a}(t) dt \right) p_{id}^{ad}(t + dt) + O(dt^2) \end{aligned} \qquad . \qquad (19)$$

Thus, the effect of the thermostatting term is a simple linear rescaling of the momentum: $p_{id}^{ad}(t + dt) \to \left( 1 - \mathbf{a}(t) dt \right) p_{id}^{ad}(t + dt) = p_{id}(t + dt)$. The moment of every particle is scaled by the same factor, $(1 - \mathbf{a}(t) dt)$. In the case of a thermostat where $m \neq 1$,

$$\lim_{dt \to 0} p_{id}(t+dt) = p_{id}^{ad}(t+dt) - \mathbf{a}(t) \left| p_{id}^{ad}(t+dt) \right|^{m-1} p_{id}^{ad}(t+dt)dt + O(dt^2)$$
$$= \left(1 - \mathbf{a}(t) \left| p_{id}^{ad}(t+dt) \right|^{m-1} dt \right) p_{id}^{ad}(t+dt) \quad (20)$$

In this case a different momentum rescaling is required for different particles and for different directions, depending upon the magnitude of the momentum in the different directions. This means that the $m \neq 1$ thermostats change the shape of the momentum distribution.

Rescaling the momentum can alternatively be considered as a rescaling of time: i.e. changing the rate of the clocks that measure the momentum evolution. For μ=1 the time rescaling is identical for all particles regardless of their momentum. However when $m \neq 1$, the time rescaling is different for different particles.

This observation has implications on the Lyapunov spectra of $m$-thermostatted systems. From the definition of the Lyapunov exponents $l_i = \lim_{t \to \infty} \frac{1}{2t} \ln \frac{d\Gamma_i^2(t)}{d\Gamma_i^2(0)}$ describing the asymptotic (exponential) rate of separation of nearby points in phase space. Non-uniform rescaling of time can be expected to result in a violation of the Conjugate Pairing Rule because fast and slow particles will be affected differently by the rescaling(s). The data presented in this paper for $m \neq 1$ thermostatted systems confirms this failure.

The failure of CPR for $m \neq 1$ thermostatted systems can also be understood by considering the structure of the stability matrix of the flow.[15] The stability matrix for the $m=1$ thermostatted systems is infinitesimally $m$-symplectic to O(1/N), however this structure is broken when a $m \neq 1$ thermostat is used. As the symmetry is broken by terms of O(1), it might be anticipated that the CPR will not be obeyed, even in the thermodynamic limit. This will be investigated in future work.

**Conclusion**

As pointed out by Klages[5] since artificial thermostatting mechanisms are models of what occurs in nature it is important to consider a range of different thermostatting mechanisms and to understand which thermostats may be used to correctly model specific systems. The present paper points out that some proposed thermostatting mechanisms have undesirable physical properties and should be used with caution. We have provided evidence for the unique status of the Gaussian isokinetic thermostat. We have shown, using a series of "*m*-thermostats" that fix the kinetic temperature

$K_2 = \sum_{i=1,d}^{N} |p_{id}|^2 / 2m$ of a system, that the Gaussian *m*=1 thermostat minimizes both the change of particle accelerations within the system and the phase space compression. While the significance of the Gaussian thermostat has been suggested previously by work on non-equilibrium systems, our work here is significant as it clearly identifies for the first time why the Gaussian isokinetic thermostat is the optimum choice for use in simulations. Indeed we show that among all *m*-thermostats, it is the *only* choice.

In this paper we have explored both equilibrium and weakly driven systems and our results clearly indicate that in the absence of a dissipative external field:

- all *m*-thermostats that violate Gauss Principle do not generate an equilibrium state and,

- among $m$-thermostats that satisfy Gauss's Principle to fix the $m+1$ moment of the velocity distribution, only the conventional Gaussian isokinetic thermostat ($m=1$) possesses an equilibrium state.

Thermostats that either violate Gauss's Principle or while obeying Gauss's Principle, attempt to constrain moments other than the second moment of the velocity distribution ($K_{m+1} = \sum_{i=1,d}^{N} |p_{id}|^{m+1}/2m$ with $m \neq 1$), result in a finite rate of phase space compression due to a continuous attempt to deform the shape velocity distribution from its canonical form. These auto dissipative thermostats fail to generate an equilibrium state. This is evidenced by the continuous compression of the accessed phase space. These results indicate that in order to permit an equilibrium state, thermostats must constrain the second moment of the velocity distribution and while so doing they must satisfy Gauss's Principle of Least Constraint. In the absence of explicit dissipative fields, such a system will on average preserve the phase space volume, The Kaplan-Yorke dimension will match the ostensible phase space dimension and the Conjugate Pairing Rule will be satisfied for adiabatically symplectic systems. For all other choices of $m$-thermostats, continuous phase space compression occurs, the Kaplan Yorke dimension will be less than the ostensible phase space dimension and the CPR cannot be satisfied.

A weak field does nothing to alter the auto-dissipative action of $m$-thermostats. Exploration of larger systems however suggests that for large N the auto-dissipative action of the thermostat is minimal and in the thermodynamic limit the properties of the system become those of a system thermostatted with a conventional Gaussian isokinetic thermostat.

**Acknowledgments**

We thank Steven Williams for useful discussions. We acknowledge financial support from the Australian Research Council and computing time provided through the Australian Partnership for Advanced Computing National Facility.

## References

[1] K. F. Gauss, J. Reine Angew. Math. **IV**, 232 (1829).

[2] W. G. Hoover, A. J. C. Ladd, and B. Moran, Phys. Rev. Lett., **48**, 1818 (1982).

[3] D. J. Evans, J. Chem. Phys., **78**, 3297 (1983).

[4] D. J. Evans, W. G. Hoover, B. H. Failor, B. Moran, and A. J. C. Ladd, Phys. Rev. A **28**, 1016 (1983).

[5] R. Klages, available online at: http://arxiv.org/abs/nlin.CD/0309069, (2003).

[6] S. R. Williams, D. J. Searles, and D. J. Evans, Phys. Rev. E, (to appear).

[7] D. J. Evans, E. G. D. Cohen, D. J. Searles, and F. Bonetto, J. Stat. Phys., **101**, 17 (2000).

[8] D. J. Evans and G. P. Morriss, *Statistical Mechanics of Nonequilibrium Liquids*. (Academic Press Limited, London, 1990).

[9] If *g* is a function of **r** and *t* only, then an additional time derivative is required to obtain an equation of form (1).

[10] D. J. Evans and A. Baranyai, Mol. Phys. **77**, 1209 (1992); and S. Sarman, D. J. Evans, and A. Baranyai, Physica A **208**, 191 (1994).

[11] W. G. Hoover, Phys. Rev. **40**, 2814 (1989);

J. Jellinek and R. S. Berry, Phys. Rev. A **40**, 2816 (1989);

Wm. G. Hoover and B. L. Holian, Phys. Lett. A, **211**, 253 (1996);

W.G. Hoover and O. Kum, Phys. Rev. E, **56**, 5517 (1997);

H.A. Posch and W.G. Hoover, Phys. Rev. E **55**, 6803 (1997) and of exponent pairs are denoted by the broken lines with open circles ($m=1$) and open squares ($m=5$). The systems consist of 4 particles at a reduced density of 0.8 and $\mathbf{F}_e = \mathbf{0}$.

**Figures**

**Figure 1:** $a$ versus $m$ for a $m$-thermostat fixing $K_2$. The system consists of 4 particles at equilibrium with a reduced density of 0.8.

**Figure 2:** $\Lambda$ (scaled in terms of the number of particles) versus $m$ for $m$-thermostats fixing $K_2$. The system size is 4 particles and the reduced density 0.8.

**Figure 3:** $a$ versus $m$ for $m$-thermostats fixing $K_{m+1}$. The system size is 4 particles and the reduced density is 0.8.

**Figure 4:** $\Lambda$ (scaled in terms of the number of particles) versus $m$ for $m$-thermostats fixing $K_{m+1}$. 4 particles were simulated at a reduced density of 0.8.

**Figure 5:** Comparison of variation of $a$ with $m$ for $m$-thermostats fixing $K_2$ in systems of differing size (4 or 50 particles). The reduced density in both systems is 0.8.

**Figure 6:** Scaled $\Lambda$ versus $m$ for $m$-thermostats fixing $K_2$ with varying system size (4 or 50 particles). The reduced density is 0.8 in both cases.

**Figure 7:** Lyapunov spectra for $m$-thermostats fixing $K_2$ with $m=1$ (filled circles) and $m=5$ (filled squares). The exponent pair index $j$ denotes pair $(l_i, l_{4N+1-i})$ and the sums

**Figure 8:** Lyapunov spectra for $m$-thermostats fixing $K_{m+1}$ with $m=1$ (filled circles) and $m=5$ (filled squares). The exponent pair index $j$ denotes pair $(l_i, l_{4N+1-i})$ and the sums of exponent pairs are denoted by the broken lines lines with open circles ($m=1$) and open squares ($m=5$). The systems consist of 4 particles at a reduced density of 0.8 and $\mathbf{F}_e = \mathbf{0}$.

**Figure 9:** $a$ versus $m$ for a series of $m$-thermostats fixing $K_2$ at varying colour field strengths. The system consists of 4 particles at reduced density 0.8.

**Figure 10:** $\Lambda$ versus $m$ for a series of $m$-thermostats fixing $K_2$ at varying colour field strengths. The system consists of 4 particles at a reduced density of 0.8.

**Figure 1**

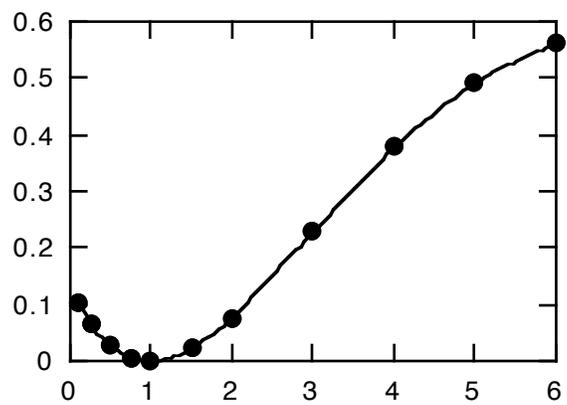

**Figure 2**

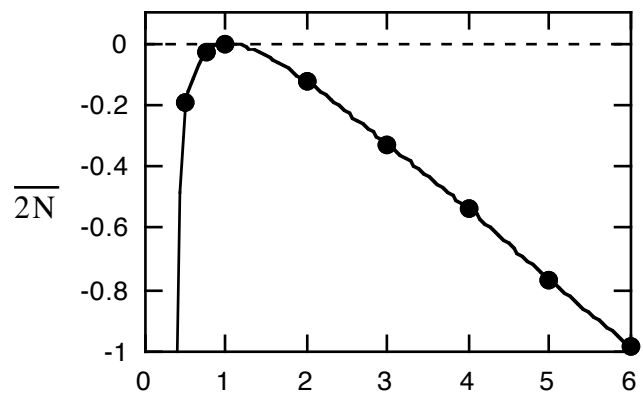

**Figure 3**

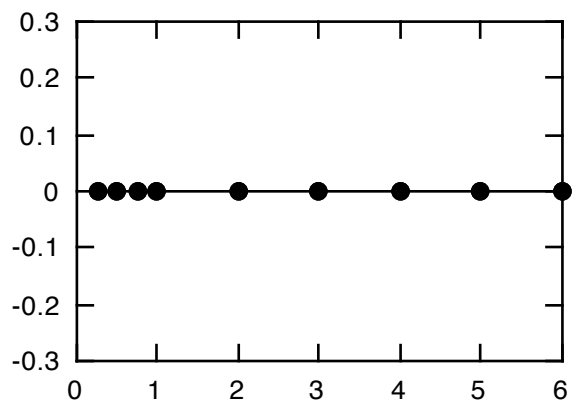

**Figure 4**

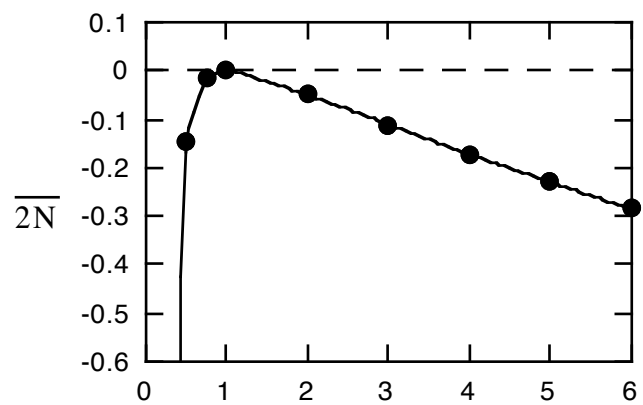

**Figure 5**

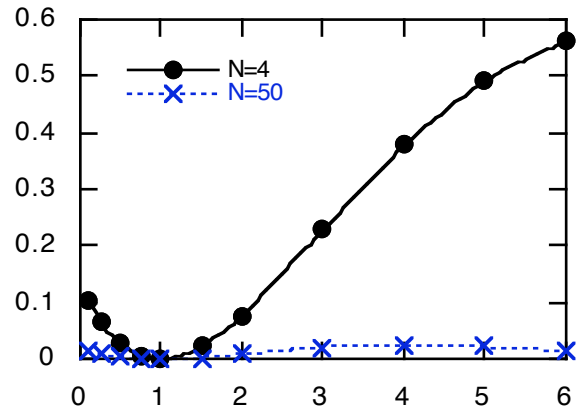

**Figure 6**

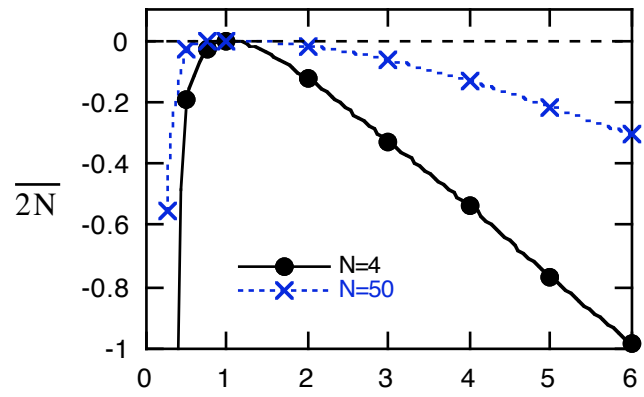

**Figure 7**

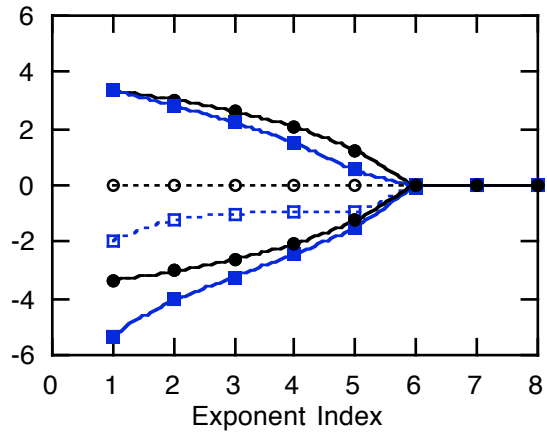

**Figure 8**

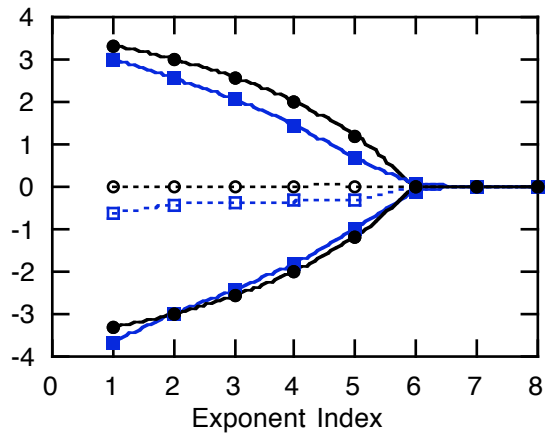

**Figure 9**

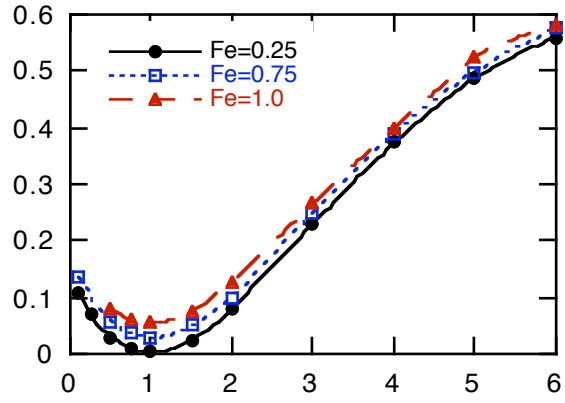

**Figure 10**

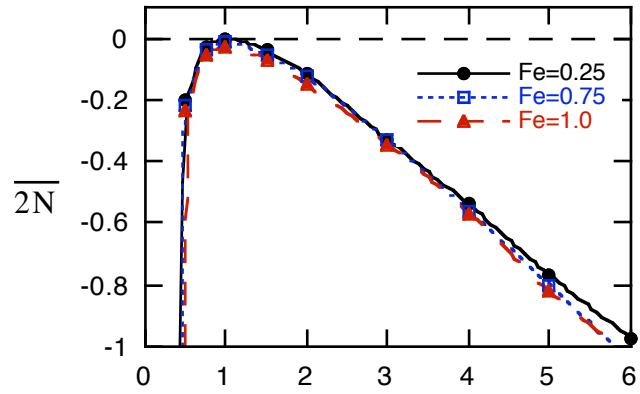